\documentclass{elsart}

\usepackage{amssymb,graphics,graphicx,epsfig}

\begin{document}

\begin{frontmatter}



\title{The Robin Hood method - a novel numerical method for electrostatic problems based on a non-local charge transfer}


\author{Predrag Lazi\'{c}},
\ead{plazic@thphys.irb.hr}
\author{Hrvoje \v Stefan\v ci\' c} 
\ead{shrvoje@thphys.irb.hr} and
\author{Hrvoje Abraham}%
\ead{ahrvoje@thphys.irb.hr}

\address{Theoretical Physics Division, Rudjer Bo\v skovi\' c Institute  \\
P.O.B. 180, HR-10002 Zagreb, Croatia}

\begin{abstract}
We introduce a novel numerical method, named the Robin Hood method,
of solving electrostatic problems. The approach of the method is  
closest to the boundary element methods, although significant conceptual differences exist with respect to this class of methods.   
The method achieves equipotentiality of conducting surfaces by iterative non-local 
charge transfer. For each of the conducting surfaces non-local charge transfers
are performed between surface elements which differ the most from the targeted
equipotentiality of the surface. The method is tested against analytical
solutions and its wide range of application is demonstrated.
The method has appealing technical characteristics. For the problem with
$N$ surface elements, the computational complexity of the method essentially
scales with $N^{\alpha}$, where $\alpha < 2$, 
the required computer memory scales with $N$, while the
error of the potential decreases exponentially with the number of iterations for many orders of magnitude of the error, without the presence of the Critical Slowing Down. The Robin Hood method has a large potential of application in other classical as well as quantum
problems. Some possible applications outside electrostatics are outlined.
\end{abstract}

\end{frontmatter}

\section{Introduction}

\label{intro}

The methods of solving electrostatic problems today range from the 
fundamentals of classical physics \cite{Jackson,Landau,Feynman,Purcell}, 
practically representing the scientific heritage, to the
state-of-the-art computational approaches \cite{Beck1,pFFT,Kinezi,ReedKorvink} 
widely used for handling complex 
%cutting-edge
technological problems. The development of computational power in
the last decades has resulted in a large number of 
available efficient methods for the
solution of potential problems in electrostatics, as well as in other areas of
physical, engineering and technological applications. According to the basic
approach towards the main goal of electrostatic problems, the determination of
the electric potential in the relevant segment of space, methods can be roughly
divided into Finite Element Methods (FEM), Boundary Element Methods (BEM) and 
Finite Difference (FD) computational methods \cite{BEMLAP,Poljak}. In very
general terms, FEM and BEM solve for the charge distribution at relevant
objects (or their boundaries) and thus obtain
the potential indirectly, whereas FD methods determine the potential directly
in the relevant segment of space.   
%the first group comprising those methods that directly
%solve the Laplace or Poisson equation in the spatial segment of
%interest; the second group consists of methods solving for the charge
%distributions at the boundaries of the relevant spatial segment, indirectly
%making possible the calculation of potential in any relevant point of space.
%The most prominent representatives of the first group are known as FEM 
%\cite{Koreanci,jos1,jos2}, while the second group is usually referred to 
%as BEM \cite{Koreanci,jos1,jos2}.

In this paper we develop a robust new method for solving large classes of 
electrostatic problems. The
usefulness and applicability of the method with respect to other potential
problems will also be outlined. 

{\bf ``As simple as it gets".} 
In this paragraph we give a clear overview of the simple
physical idea that has guided us towards the development of the method 
presented here.
Owing to the abundance of technical details given later, the main idea could be
blurred and that simple idea is what
gives our method efficiency and robustness.
To give the insight into the core of the method, we describe a simple problem. 
Suppose that we have an ideal, insulated and charge neutral, 
metal sphere standing in vacuum and we bring a point charge next to it.
What will happen with the charge distribution on the sphere? It will 
redistribute until all the surface of the sphere becomes equipotential. 
That is the stationary situation: the charge will not redistribute any 
further because the potential is the same everywhere on the sphere 
surface. That is a very simple high-school argument which reveals the qualitative 
nature of the stationary solution. With equal simplicity, one can 
deduce that the electric field must be perpendicular to the surface of the metal,
as indeed will be seen in the examples treated in this paper. 
Following only this equipotentiality principle, we employ a straightforward 
numerical procedure to find a complete quantitative solution of this 
electrostatic problem. First, we divide a sphere into finite triangle elements, each having some surface charge. The initial surface charge distribution is chosen in such a way to respect the charge neutrality of the sphere. One can simply set the charge distribution to zero on the entire sphere. We calculate the electrical 
potential at each of these elements due to charges on all triangles and a point
 charge.
We determine two of the triangles which have the highest and the lowest potential, 
respectively,
%(negdje treba spomenuti da nije bitno ako ih ima vise jednakih maximuma ili 
%minumuma tj. degeneracija nije nikakav problem). 
and transfer the charge 
from one of these two selected elements to the other in such a manner that 
after the transfer the potentials on these two elements are exactly the same.
Since we do only the charge transfer, the total charge on the sphere remains 
conserved, i.e. the sphere remains neutral. Then the update of the potential, 
resulting from the charge transfer, is performed. 
We iterate this process. The main idea is that such a procedure will lead to 
a more and more equipotential surface, and eventually converge to the solution of the 
electrostatic problem. 
%It is intuitively clear that such a procedure could lead to a
%solution, 
%but we could not provide a strict mathematical proof.  
Since the main idea of the method, namely taking from the maximum and giving to the 
minimum thus making them equal, to the principal ideas of Robin Hood (RH),  we 
suggest this name for out method.

The conceptual importance of this ``as simple as it gets" reflection is 
the reason for placing this brief description of the method already here in the
introduction. All properties of the RH method stem from this main 
simple idea of min-max equipotentialization and {\em not} from the particular
implementation. As a matter of fact, 
%one could stop reading this paper here, because our main idea is already described, one 
% can immediately start playing with it
all essential elements of the RH method are provided in the description 
given above, which best illustrates the conceptual attractiveness of the RH method.
The rest of the paper is devoted to the elaboration of the stated principles and 
the description of the elements of implementation of the RH method which raise 
afore-mentioned simple ideas to the level of a powerful calculational technique.    

%The presentation of the method is arranged in such a
%manner as to provide the best insight into the physical motivation behind the
%method itself. 
The paper is organized as follows: 
The second section is devoted to the physical foundations of the RH method. 
This section gives general elements of the method as well as the specific
features for the cases of insulated conducting surfaces and conducting surfaces
at the exterior potential. The third section specifies details of the
implementation of the method. The fourth section comprises several examples, some of
which show the reliability of the RH method by comparison with analytical
solutions, while others demonstrate the broad applicability of the method.
The fifth section exposes the technical characteristics, like the computational
complexity, memory requirements, speed of convergence and others. The sixth
section outlines the possibilities of extension of the RH method beyond
electrostatics. The seventh section closes the paper with conclusions.

\section{Physical foundations} 

\label{phys}

A large class 
%\footnote{There are some theoretically
%interesting configurations of electric fields without conductors \cite{Jackson}.} 
of electric field configurations is achieved by an appropriate
spatial configuration of various conductors. These conductors, such as
electrodes, cables, plates, etc. are either maintained at the potentials of
exterior voltage sources or are insulated. For static electric fields, there are no
electric currents and all parts of every conductor are at the same
potential. The principle of equipotentiality of conductors in electrostatics is
at the very core of the computational method presented in this paper. 

Despite the fact that here we principally consider the electrostatic case, it
is instructive to take a look into the process of attaining the static
configuration of the electric field. When one of the conductors is attached
to an exterior voltage source or charges are deposited close to the surfaces of
the conductors, the electric currents flow and rearrange the distribution of free
charges in the conductors until the surfaces of the conductors become
equipotential again. Therefore, every static configuration of the electric field is
initially achieved by the redistribution of the free charge in the conductors.
The description of the process of reaching some electrostatic configuration is 
generally more complex than the description of the electrostatic configuration
itself. However, it is tempting to investigate the usefulness of 
the concept of the charge
redistribution in determining the electrostatic configurations
themselves. Here we present the method of solving electrostatic problems using 
the iterative redistribution of the charge at the surfaces of conductors. 

We consider the general problem of determining the electric potential in the
spatial segment delimited by conducting surfaces. The underlying assumption
of the method is that the Green function of the system is known, i.e. that with
the knowledge of the charge distribution at all surfaces it is possible to
determine the electric potential at any point of interest. This assumption
represents the most fundamental limitation of the method, but is, on the other
hand, justified for a very large class of both theoretically and practically
interesting problems. The conducting surfaces
either are kept at defined potentials or are insulated. Every conducting surface is
divided into surface elements and a point is chosen within each surface element.
The potential is calculated in the chosen point and in the remainder of the 
text this point will be referred to as the {\em point of calculation} (POC).
The area of the surface element $i$ is denoted by $\Delta S_{i}$, the
coordinates of its POC by $\vec{x_{i}}$ and the potential at its POC by $U_{i}$.  
In the practical implementation of the method, described in section
\ref{impl}, the 
surfaces are divided into triangles and the POCs are the barycentres
 of the triangles. This choice is motivated
 by the fact that the dipole of the uniformly charged triangle vanishes when it is
 calculated in the reference frame centered at its barycentre. This feature will be
 exploited in the implementation of the RH method. The aim of the method is
to achieve the equipotentiality of each of the conducting surfaces.
Practically, this is realized by achieving the equality of the potential at all
POCs for every individual conducting surface, to a predefined accuracy. In the
case of conducting surfaces kept at a fixed potential, 
the potential achieved at POCs
is the potential of the exterior voltage source, while for the
insulated conducting surfaces, the achieved potential is obtained as one of the results
of the method. It is assumed that the surface charge density is constant within
every surface element. The surface charge density of the surface element $i$ is
$\sigma_{i}$, while its charge is $q_{i} = \sigma_{i} \Delta S_{i}$. 

In the process of finding the static distribution of the charge at the conducting
surfaces, we start from some initial distribution of the charge at each of the
conducting surfaces. Using the known Green function, it possible to calculate
the potential at each POC coming from the entire charge distribution in the system:
\begin{equation}
\label{eq:potential}
U(\vec{x}) = \int_{\partial V} G(\vec{x}, \vec{x'}) \sigma(\vec{x'}) \, dS' \; ,
\end{equation}
with the Green function $G(\vec{x}, \vec{x'}) = 
%(4\pi \epsilon_{0} 
k \mid \vec{x}-\vec{x'}\mid^{-1}$ and $\partial V$ stands for all boundary 
surfaces of the system. In
our method, the analytic expression (\ref{eq:potential}) 
can be cast into the discretized form 
\begin{equation}
\label{eq:potdiscrete}
U_{i} = \sum_{j} I_{ij} q_{j}\; ,
\end{equation}
where
\begin{equation}
\label{eq:Iij} 
I_{ij}= \frac{1}{\Delta S_{j}} \int_{\Delta S_{j}}
%(4\pi \epsilon_{0} 
k \mid \vec{x_{i}}-\vec{x'} \mid^{-1} \; dS' \, .
\end{equation}
In the calculation of the potential at a given POC one needs to compute the
contributions of all surface elements, including the contribution from the
surface element within which the POC is situated. Once the potential at all
POCs is known, we can find a pair of POCs at which potential differs the most 
from the target value. Finally, charge transfers to/from chosen surface elements
(where POCs from the chosen pair are situated) are performed in order to
equalize the potentials at the POCs of the chosen pair. The procedure of
equalizing potential differs somewhat, depending on whether the conducting surface
is kept at the exterior potential or is insulated. Therefore we describe this
procedure separately for each of the two cases.

{\bf Insulated conducting surface.} In this case, 
there is no target potential which needs to be achieved in the process
of determining the static surface charge distribution. The only condition
that must be satisfied is the equipotentiality of the surface. Therefore, we
choose two points which differ the most from the targeted equipotential
configuration: the POC of the maximal potential and the POC of the minimal
potential at the surface. We denote the POCs of the maximal and the minimal potential by
$m$ and $n$, respectively. The potentials of the two chosen POCs are
equalized by the charge transfer from the point of the maximal potential to the
point of the minimal potential. If the amount of charge $q$ is transferred from $m$
to $n$, the potentials at these two points after the transfer will be
\begin{eqnarray}
\label{eq:newpot}
U_{m}' &=& U_{m} - I_{mm}q + I_{mn}q \; , \nonumber  \\
U_{n}' &=& U_{n} + I_{nn}q - I_{nm}q \; . 
\end{eqnarray}
The condition of equality of the potential at the points $m$ and $n$ after the
charge transfer ($U_{m}'=U_{n}'$) yields the amount of charge to be transferred:
\begin{equation}
\label{eq:charge}
q=\frac{U_{m}-U_{n}}{I_{mm}+I_{nn}-I_{mn}-I_{nm}} \, .
\end{equation}
The transfer of the charge (\ref{eq:charge}) clearly influences the
potential values at other POCs and generally brings a large majority of 
them closer to the requirement of
equipotentiality. This form of charge transfer also ensures the crucial property
of the conservation of charge. In the case of multiple insulated conducting
surfaces in the system, charge transfers are performed separately within 
each surface. Charges at other surfaces, however, influence the potentials at
any individual surface.

{\bf Conducting surface at the exterior potential.} For a conducting surface at the
exterior potential $U_{ext}$, the targeted value of the potential is defined. Two POCs,
denoted by $k$ and $l$, at which the potential differs the most from the external
potential value, are located. The amounts of charges $q_{k}'$ and $q_{l}'$ are brought
to the POCs $k$ and $l$, respectively. The values of these charges are
determined from the condition that the potential at the chosen POCs after
introducing these charges should equal the potential of the external voltage source.  
After the charges $q_{k}'$ and $q_{l}'$ are transferred to the points $k$ and $l$, the
potentials at these two POCs are
\begin{eqnarray}
\label{eq:newpotfixed}
U_{k}' &=& U_{k} + I_{kk}q_{k}' + I_{kl}q_{l}' \, , \nonumber  \\
U_{l}' &=& U_{l} + I_{ll}q_{l}' + I_{lk}q_{k}' \, . 
\end{eqnarray} 
The right amount of charges to bring the potentials at $k$ and $l$ to the external
potential ($U_{k}'=U_{l}'=U_{ext}$) are obtained by solving the system of linear equations
(\ref{eq:newpotfixed})
\begin{eqnarray}
\label{eq:charges2}
q_{k}' & = &
\frac{(U_{\mathrm{ext}}-U_{k})I_{ll}-(U_{\mathrm{ext}}-U_{l})I_{kl}}{I_{kk}I_{ll}-I_{kl}I_{lk}}
\, , \nonumber \\
q_{l}' & = & \frac{(U_{\mathrm{ext}}-U_{l})I_{kk}-(U_{\mathrm{ext}}-U_{k})I_{lk}}
{I_{kk}I_{ll}-I_{kl}I_{lk}} \, .
\end{eqnarray}
The sum of transferred charges $q_{k}'$ and $q_{l}'$ is generally not zero
because
there is the possibility of charge flow to/from the exterior voltage source.  

Both procedures described above correct the potential at points where 
the greatest deviations from the targeted equipotentiality exist. 
%This fact probably
%contributes the most to the fast exponential convergence of the procedure. 
The entire procedure is iterated for all conducting surfaces until the 
criterion of equipotentiality is satisfied for each surface.  

In both cases presented above it is possible that at some iteration multiple
minima or maxima exist (or equivalently, points with the largest departure from the
exterior potential). This kind of situation is most likely to appear at
the beginning of the iteration process when the system possesses some
symmetry. This possibility of degeneration, however, presents no 
difficulty for the RH
method. It is sufficient to choose one of the maxima and one of the minima
and to proceed with the algorithm of the method. 
For both cases it is also possible to develop procedures in which an 
arbitrary number of points $N_{\mathrm{points}}$ is brought to the targeted 
potential at a single step of iteration. Such procedures require
finding the solution of the $N_{\mathrm{points}}$ dimensional set of linear equations.
For the conducting surface at an exterior potential, the choice of a single 
point ($N_{\mathrm{points}}=1$) is also allowed in principle. However, in this case, the 
stability of computations becomes questionable. Furthermore, it is our
view that the procedure with $N_{\mathrm{points}}=1$ fails to exploit the 
possibilities of fast convergence since it completely ignores the typical 
scales of the charge distribution at the studied surface. The question of
quality of convergence for $N_{\mathrm{points}}$ larger than 2 is presently open. 
%and will be addressed elsewhere.
For $N_{\mathrm{points}}$ very large, i.e. close to the number of surface elements, the
method conceptually approaches the existing BEM. For the time being, we find the choice 
$N_{\mathrm{points}}=2$ to be conceptually the simplest and adopt it in the remainder of
the paper. Furthermore, in the extensions of the RH method to the problems in 
which the potential depends non-linearly on the density (e.g. Thomas-Fermi model),
or where the potential at some POC is dependent on densities of surrounding POCs
(e.g. in representations of terms with derivatives) or densities are strictly non-negative
(e.g. in quantum problems), the case of $N_{\mathrm{points}}$ larger that 2 becomes 
increasingly cumbersome.

\section{Implementation of the method}

\label{impl}

The focus of this section is on the practical implementation of the procedures explained
in the preceding section. There are several major segments of implementation
which are combined together into a robust and efficient calculational scheme.
Each of these segments also introduces its discretizational error which can be
reduced by discretization refinement. 
The first segment of implementation is the division of surfaces into surface
elements $i$. In our method, all surfaces are divided into triangles using 
algorithms included in {\em VTK} \cite{VTK} and {\em Mathematica}
\cite{Math} program packages. Clearly, this step introduces the error of 
approximating a (generally piece-wise smooth) 
surface with a set of triangles. This error can be reduced by
increasing the number of triangles used to approximate the studied surface. Each
of the triangles is then divided into two right-angled triangles, which can be
carried out in at least one way. In this way we approximate the surface by a
set of right-angled triangles. The choice of right-angled triangles is not fundamental, but is motivated by practical and numerical reasons. 
As specified in the preceding section, we take that
the surface charge density is uniform within each of the triangles. For each
right-angled triangle its POC is situated at its barycentre.
Once the set of approximating right-angled triangles is defined, the integrals
$I_{ij}$ (\ref{eq:potdiscrete}) need to be calculated. For the right-angled
triangle $i$ with the catheti $a_{i}$ and $b_{i}$ the value of its {\em
self-contribution} $I_{ii}$ can be obtained analytically:
\begin{eqnarray}
\label{self}
I_{ii} & = & k a_{i} \left[ \ln \left[ (2s_{i}+t_{i,1})/(t_{i,3}-s_{i}) 
\right] \right. \nonumber \\
&+& s_{i} \left[ \ln \left[ (t_{i,4}+t_{i,2} t_{i,3}+2)/(t_{i,1} t_{i,3}-2 t_{i,4}-1)
\right] /t_{i,3} \right.
\nonumber \\
&+&  \left. \left. \ln \left[ (3 t_{i,2}+6)/(3 t_{i,3}-3) \right] \right]
\right]/3.
\end{eqnarray}
Here we have used abbreviations $s_{i}=b_{i}/a_{i}$, $t_{i,1}=\sqrt{1+4 s_{i}^{2}}$,
$t_{i,2}=\sqrt{4+s_{i}^{2}}$ and $t_{i,3}=\sqrt{1+s_{i}^{2}}$.
For the general integral $I_{ij}$, one must, however, apply non-analytic
procedures. One option is the numerical calculation of the integral. This is
generally rather time-consuming and will be avoided.
% \footnote{
%Numerical calculation of
%$I_{ij}$ might represent especially large drawback in the case of the variant of
%our method that requires memory space proportional to the number of right-angled 
%triangles.}. 
The other possibility is the expansion of the integral in a
multipole expansion \cite{Jackson}. In a multipole expansion for the potential
of a given charge distribution we can  increase the quality of the approximation 
by adding the contributions of the multipoles of higher rank. The
calculation of higher-rank multipoles, however, becomes increasingly
cumbersome. On the other hand, for a calculation up to some fixed multipole 
rank, improvement in precision can be achieved by dividing the initial
charge distribution into several smaller charge distributions. 
In this approach, however, the amount of calculation grows with more
detailed divisions of the initial charge distribution. We find that it is
optimal to use the combination of these two approaches. Therefore, in our
calculations we combine the use of multipole moments up to the quadrupole and apply the recursive division of the initial right-angled triangle into smaller right-angled triangles.
Let us consider this combination in more detail. 

For a right-angled triangle in the
$x-y$ plane with the catheti $a_{j}$ and $b_{j}$ oriented in the $x$ and $y$ 
axes, respectively,
the dipole moment calculated with respect to its POC vanishes
(since for the POC of the triangle we choose its barycenter). 
The quadrupole 
moment in the reference frame of POC is given by     

\begin{equation}
Q = \frac{2}{a_{j}b_{j}}\left[ \matrix{ \frac{a_{j}^3\,b_{j}}{18} - 
\frac{a_{j}\,b_{j}^3}{36} & 
-\frac{ a_{j}^2\,b_{j}^2 }{24} & 0 \cr -\frac{ a_{j}^2\,b_{j}^2  }
   {24} & -\frac{ a_{j}^3\,b_{j}  }{36} + \frac{a_{j}\,b_{j}^3}{18} & 0 \cr 0 
   & 0 & -\frac{ a_{j}^3\,b_{j}  }{36} - 
   \frac{a_{j}\,b_{j}^3}{36} \cr  } \right] \, .
\end{equation}

%\begin{equation}
%\label{eq:quad}
%Q= \frac{2}{a_{j}b_{j}}\left[ \begin{matrix}
% \frac{a^3\,b}{18} - \frac{a\,b^3}{36} & - \frac{a^2\,b^2}{24} & 0 \cr 
% -\frac{a^2\,b^2}{24} &  -\frac{a^3\,b}{36} + \frac{a\,b^3}{18} & 0 \cr 
% 0 & 0 &  -\frac{a^3\,b}{36} - \frac{a\,b^3}{36} \cr   
%\end{matrix} \right]        
%\end{equation}

\begin{figure}
\centerline{\rotatebox{-90} {\resizebox{0.25\textwidth}{!}
{\includegraphics{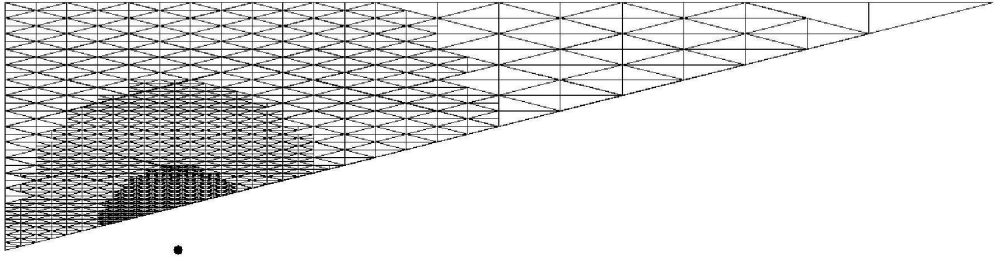}}}}
\caption{\label{fig:rectriangle} The result of the iterative division of the 
right-angled triangle with the vertices A=(1,0,0),
B=(0,4,0), C=(0,0,0) for the computation of the potential at the point
T=(1,0.7,0), denoted by the heavy dot.}
\end{figure}

The integral $I_{ij}$ of the right-angled triangle up to the quadrupole 
contribution at the point $\vec{x_{i}}$ is given by the expression

\begin{equation}
\label{eq:potquad}
I_{ij}=
%\frac{1}{4 \pi \epsilon_{0}} 
k \left[ \frac{1}{\mid \vec{x_{i}} -
\vec{x_{j}} \mid} + \frac{1}{6}  Q_{mn} \frac{
(\vec{x_{i}} -\vec{x_{j}})_{m} (\vec{x_{i}} -\vec{x_{j}})_{n}}
{\mid \vec{x_{i}} - \vec{x_{j}} \mid^{5}} \right] \, .
\end{equation}

The expression given above is applicable for surface elements $i$ and $j$
sufficiently far apart. However, if these elements are close, this approximation
may not suffice for some predefined accuracy in the calculation of the potential.
The criterion for the applicability of the expression (\ref{eq:potquad}) is that
the ratio of the distance $\mid \vec{x_{i}} - \vec{x_{j}} \mid$ to the typical
size of the right-angled triangle is larger than some number. The measure of the
size of the right-angled triangle is taken to be the larger of its catheti. 
According to our numerical analysis,
when the value of the ratio of $\mid \vec{x_{i}} - \vec{x_{j}} \mid$ and the
larger cathetus (let us denote it by $\eta$) is $ \ge 5.5$, 
the quadrupole corrected monopole term (\ref{eq:potquad}) gives 
the approximation of the result obtained by the direct numerical integration
with an accuracy of $10^{-4}$. Furthermore, when the ratio $\eta$ is $\ge 26$,
the monopole term alone is sufficient to provide the approximation with 
the same
accuracy. These findings are valid for all right-angled triangle shapes, i.e. for
all ratios of the larger and smaller catheti. Therefore, if the ratio $\eta$
is $ \ge 26$, the monopole term is used and if the afore-mentioned ratio is 
$\ge 5.5$, the
monopole plus quadrupole terms are used. If, however, we have $\eta < 5.5$, the
right-angled triangle is divided into four similar right-angled triangles obtained
by the bisections of the sides of the original right-angled triangle. It is
important to note that the respective $\eta$ ratios of the four newly formed
right-angled triangles are generally larger than those of the original triangle.
The procedure is then repeated for each of the four right-angled triangles and
subsequently iterated until all obtained right-angled triangles satisfy the
condition $\eta \ge 5.5$. The result of one such iterative division for an elongated
right-angled triangle and a nearby point is shown in Fig. \ref{fig:rectriangle}.
Summation of contributions of all triangles yields the final result. At this place it is important to stress that the subdivision of the right-angled triangle depicted in Fig.
\ref{fig:rectriangle} is not a rafinement of the discretization of the surface, but only an auxiliary tool during the calculation of the potential. The
combination of the multipole expansion up to the quadrupole and the iterative division
of the triangle surface provides a fast, reliable and robust method for the
calculation of $I_{ij}$ quantities. 

During the entire execution of the RH method only the potentials at all POCs 
are stored. The quantities $I_{ij}$ are calculated {\em over again} at each
instance when they are needed, i.e. they are {\em not recorded at any
instant}
during the execution of the algorithm. If the algorithm of the method should
include storing of the quantities $I_{ij}$, memory requirements would
grow quadratically with the number of surface elements $N$. In this case, an
advantage would be that these quantities would have to be calculated only
once. However, we do not adopt this approach. Since only potentials are
recorded in the RH method, memory requirements grow {\em linearly} with $N$.
This feature is clearly a large advantage. The disadvantage lies in the
necessity of a new calculation of the quantities $I_{ij}$ whenever they are
needed. However, owing to the essential characteristics of the RH method, this
disadvantage is not so severe. Namely, the algorithm of the RH method can be
divided into two stages. In the first initializational stage, the values of the potential
are calculated at all POCs from some initial charge distribution. In this
stage the number of required calculations of the quantities $I_{ij}$ grows as
$N^2$ (it is important to note that the times necessary for all these 
calculations are not the same). In the second stage, the iterative non-local
charge transfers are made. A very important characteristic of the RH method
emerges in this stage. The potentials at all POCs after any iteration
(charge transfer) need not be completely calculated, but {\em only updated}.
Namely, it is only necessary to add changes due to the charge transferred between
the two chosen POCs. In this update the number of required calculations of
$I_{ij}$ scales as $N$. In this way, the potential is largely reused from
one iteration to the other. This feature of the potential updating is at the
core of the many powerful characteristics of the RH method.
Recalculation of $I_{ij}$ at every step fully justifies great
attention that was paid to the choice of the optimal method for the calculation
of these quantities, which was elaborated in the preceding paragraphs. The
choice of the optimal calculational method for $I_{ij}$, together with the
updating feature described above, makes the execution times of the RH method
acceptable. 

The RH method has also attractive properties with respect to parallelization.
In the parallelized version of the RH method, each of the processors would
be assigned a subset of surface elements. The advantages of
parallelization can be seen, e.g. in the update of the potential. 
Once in some iteration the charge transfer is performed, each of the
processors simultaneously calculates the update of the potential in its
realm. The possibility of efficient parallelization also significantly
improves the execution time of the RH method.
     
Another rafinement of the method is available if we consider the dynamical
approach to the division of the studied surface into triangles. In the present paper, the
initial division depends only on the surface geometry, while it is completely
insensitive to physical quantities. In such a setting, it is possible that an
approximation of some conducting surface even with a large number of triangles
may in some areas where physical quantities have large gradients be inadequate 
and locally provide results of lower quality. Clearly, the solution of this problem
is in dividing the existing triangles into smaller ones in the areas where the
refinement is needed. Within our method,
it is especially easy to accommodate this form of
refinement of the approximation of the studied surface with a set of right-angled
triangles. There are two principal reasons for this. The solution
satisfying the condition of equipotentiality for the cruder set of triangles
can be used as a starting distribution for the iteration with the refined set of
triangles. This further accelerates the computation. Furthermore, it is
straightforward to specify criteria which determine which triangles should be
further divided. Thus the iteration procedure naturally leads to the
determination of areas where the refinement is required. One criterion might be
that the percentage $\alpha$ of all triangles with the largest charges are
further divided. Another criterion might be to consider the average or overall 
difference of the potential at the vertices of the triangle and its POC. For
example, one might consider the quantity $G_{i} = (U(\vec{x_{i}}) - 
U(\vec{x_{i}}^{A}))^2+(U(\vec{x_{i}}) -U(\vec{x_{i}}^{B}))^2 +
(U(\vec{x_{i}}) -U(\vec{x_{i}}^{C}))^2$, where $\vec{x_{i}}$ denotes the POC of
the triangle $i$, while $\vec{x_{i}}^{A,B,C}$ stand for its vertices. In this
case, further division would be performed on those triangles for which 
$G_{i}^{1/2}/U(\vec{x_{i}}) \ge \beta$, where $\beta$ is a predefined factor.
The entire computational scheme would then include iterations of a composite
procedure: for a given set of triangles, the equipotentiality at their POC would
be achieved and then, using one of the criteria specified above, the set of
triangles would be refined by the division of some of them. The application of
this composite iterative procedure yields a powerful computational method for
interesting classes of electrostatic (and other) problems.    
             
The adaptive subdivision, as described in the preceding paragraph, refines only the set of triangles obtained in the initial discretization of the surface. 
%A possible advantage of such an approach is in its modularity.
The vertices of the new triangles lie in the planes of the old triangle, not on the original surface , which is in general case curved. 
The procedure that discretizes the surface just produces the set of triangles, and the RH method uses these triangles during the entire execution and possible adaptive subdivision. The set of triangles is an input for the RH method. The full adaptive approach would have to include the refinement of the surface discretization and not only the subdivision of the initial discretization of the surface. In such an approach it becomes necessary to integrate the discretization procedure with the RH method. This approach has not been pursued in this paper that serves as an introduction to the RH method. For the high precision technical applications, such as e.g. investigation of ``hot spots", this approach needs to be adopted. 
%In this way, the modularity property would be lost, but the method would become suitable for the %calculations with great precision. This modification of the RH method is a path worthy %pursuing.   

\section{Examples}

\label{ex}

The main goal of this section is to display solutions obtained using the RH method.
%The strength and scope of any computational method can be best assessed through
%numerous and diverse examples.
For each characteristic of the method we specify
examples demonstrating its advantages. In some of the examples given below, 
the units of dimensional quantities are left unspecified since the results
are valid for arbitrary units.  

\subsection{Comparison with analytic results}

\label{anal} 

\begin{figure}[h]
\centering
\includegraphics[clip=true,scale=0.5]{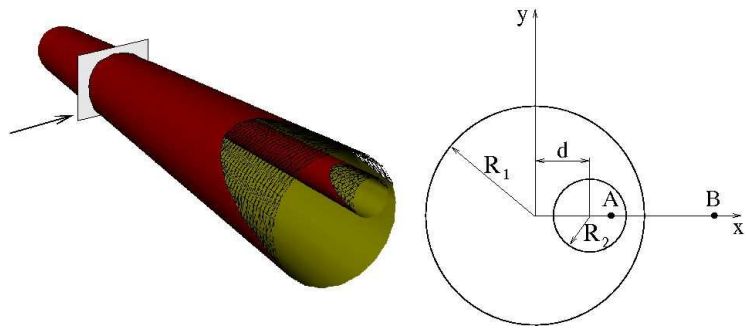}
\caption{\label{fig:cil} The geometrical configuration of the cylinders forming the capacitor. The arrow in the left picture shows the slice for which the potential and the electric field are calculated and shown in Fig. \ref{fig:field}}
\end{figure}

In this subsection we discuss solutions of two problems obtained by our computational method and compare the numerical results with the analytical solutions. 
To bring up several points we choose two significantly different problems:
\begin{itemize}
\item[i)] a metal grounded sphere of radius $R$, centered at the origin and a point charge $q$ at a distance $y$ from the centre of the sphere ($y > R$)
\item[ii)] two infinite parallel metal cylinders, one completely inside the other as shown in Fig. \ref{fig:cil}. The outer cylinder is charged with charge +Q and the inner with -Q i.e. they form a capacitor.
\end{itemize}

The first problem is easily accessible by the RH numerical method due to the finite size of the objects involved in the calculation (the sphere and the point charge).
This problem is analytically solved in detail in \cite{Jackson} using the method of images. 
The induced charge on the sphere has value $q_{ind}=-qR/y$. Our solution for the parameters $y=3$, $R=2$, $q=10$ and the sphere represented by 139240 triangles yields $q_{ind}=-6.6664$ while the exact value is $-20/3=-6.666666$ which is 0.004$\%$ difference.  

The second problem of capacitance of two infinite cylinders can not be done in numerical calculation without approximation of the infinite cylinders by the finite ones. Using the finite instead of the infinite cylinders will introduce the charge distribution at the ends of the cylinders different from the one obtained in exact analytical solution for infinite cylinders. Nevertheless, one can efficiently eliminate the contribution of these boundary effects in the final solution for the capacitance per unit length. Generally, when we calculate capacitance of some system, we put charges of equal size and different sign (namely $+Q$ and $-Q$) on two objects representing the ``plates" of the capacitor and then solve for the equipotentiality of each of these isolated charged objects. By that we obtain the potential of each plate and the potential difference U between the capacitor plates. The capacity of such a system is then given by definition as $C=Q/U$.
In this case we know that the calculated system is different from the infinite system and that the main difference is in boundary effects that are present in the numerical solution. Therefore we perform two calculations with different lengths of cylinders. 
The boundary effects contribution to capacitance of the system can be canceled and we can obtain a very accurate value of $C/L$ for the infinite cylinders in the following way: if the capacitance in the calculation in which the length of the cylinders is $L_1$ is $C_1$ and the capacitance of the cylinders of length $L_2$ is $C_2$ then the capacitance of the infinite capacitor per unit length L would be 
\begin{equation} 
\label{eq:capac}
\frac{C}{L}=\frac{C_2 - C_1}{L_2 - L_1}.
\end{equation} 

One can solve this problem analytically by the method of images replacing two cylinders with two homogeneously charged wires at positions A and B as shown in Fig. \ref{fig:cil} and require for a constant potential at each cylindrical surface. The obtained solution gives capacitance per unit length
\begin{equation}
\label{eq:ancapac}
\hspace{-1cm}
\left(\frac{C}{L}\right)^{-1}=2 \ln \left(\frac{R_1^2+R_2^2-d^2+\sqrt{R_1^4+R_2^4+d^4-2d^2R_1^2-2d^2R_2^2-2R_1^2R_2^2}}{2R_1R_2} \right)
\end{equation}

The two calculations were performed with the following parameters:
i) for $d=0.25$, $R_1=1.0$, $R_2=0.25$, $L_1=15.6$ and both cylinders represented by N=75000 triangles, the obtained capacitance per unit length is $C_1/L_1$=0.391198.
ii) for $d=0.25$, $R_1=1.0$, $R_2=0.25$, $L_1=19.6$ and both cylinders represented by N=75000 triangles, the obtained capacitance per unit length is $C_2/L_2$=0.394214.

When we calculate $C/L$ as given in equation (\ref{eq:capac}), we obtain $C/L=0.379433$ which is only $0.06 \%$ different from the analytical value.

\begin{figure}[h]
\centering
\includegraphics[clip=true,scale=0.5]{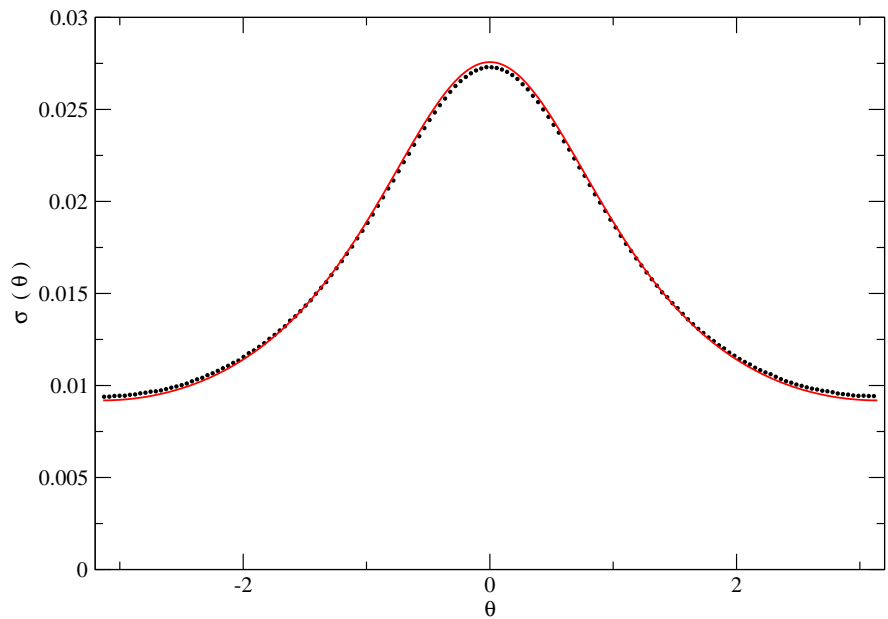}
\caption{\label{fig:cilpot} The angular dependence of the surface charge density of the outer cylinder. The very good agreement of the numerical and the analytical curve is evident.}
\end{figure}

We shall exploit this example further to demonstrate how the RH method finds the distribution of surface charge in such a system which possesses the mirror plane symmetry (it can also have axial symmetry in case of concentric cylinders i.e. when $d=0$).
In the analytical solution one can find the electric potential and the electric field everywhere in the space and particularly at the surface of the cylinders. From the value of the electric field at the cylinder surfaces one can find the surface charge on them. We compare the surface charge of the outer cylinder found by the RH method with the values from the analytical solution. Due to the boundary effects, we choose for comparison one slice from the middle of the cylinder as shown in Fig. \ref{fig:cil}. 
In Fig. \ref{fig:cilpot} is given the comparison between the numerical and the analytical solution for the surface charge. The agreement is very good, considering that the analytical solution is obtained for infinite cylinders.
To show how the RH solution respects the symmetry of the system, we calculate the electric potential and the electric field in the plane perpendicular to the axes of cylinders and passing through the middle of their length. 
The symmetry of the RH solution perfectly respects the symmetry of the problem even though in the RH method information on symmetry on the problem is not imposed in any way and enters the calculation process only through the positions of input triangles.

\vspace{1cm}

\begin{figure}[h]
\centering
\includegraphics[clip=true,scale=8.0]{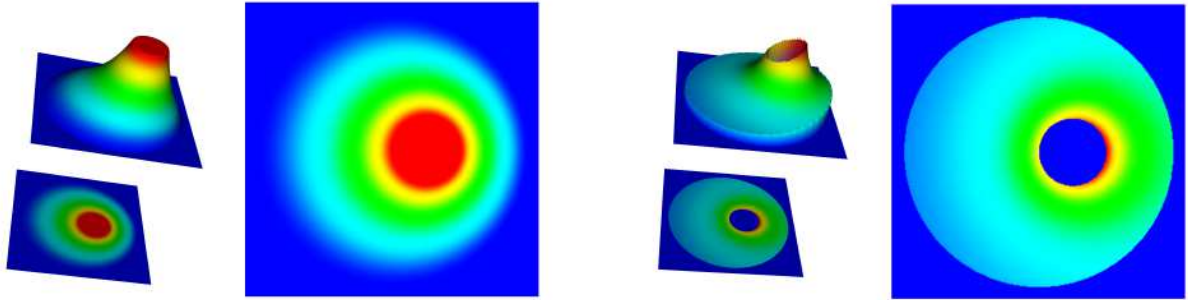}
\caption{\label{fig:field} Once the distribution of the surface charge is found by the RH method, one can calculate the electric potential (three figures on the left) and the electric field (three figures on the right) in the whole space which is very important in the study of discharge, ``hot spots", etc. In the set of pictures describing the electric field, the absolute value of the electric field vector is shown. The calculations were done for the slice shown in Fig. \ref{fig:cil}} 
\end{figure}

\subsection{Potential problems in complex configurations and geometries} 

\label{compl}

In this subsection we present solutions of several potential problems with
complex configurations and geometries. These examples are designed to 
illustrate that the numerical
method introduced in this paper can quite easily handle even very demanding
geometries and configurations, i.e. shapes and relative positions of the 
conducting surfaces. 

\begin{figure}
\centerline{%\rotatebox{-90}
{\resizebox{1.0\textwidth}{!}
{\includegraphics{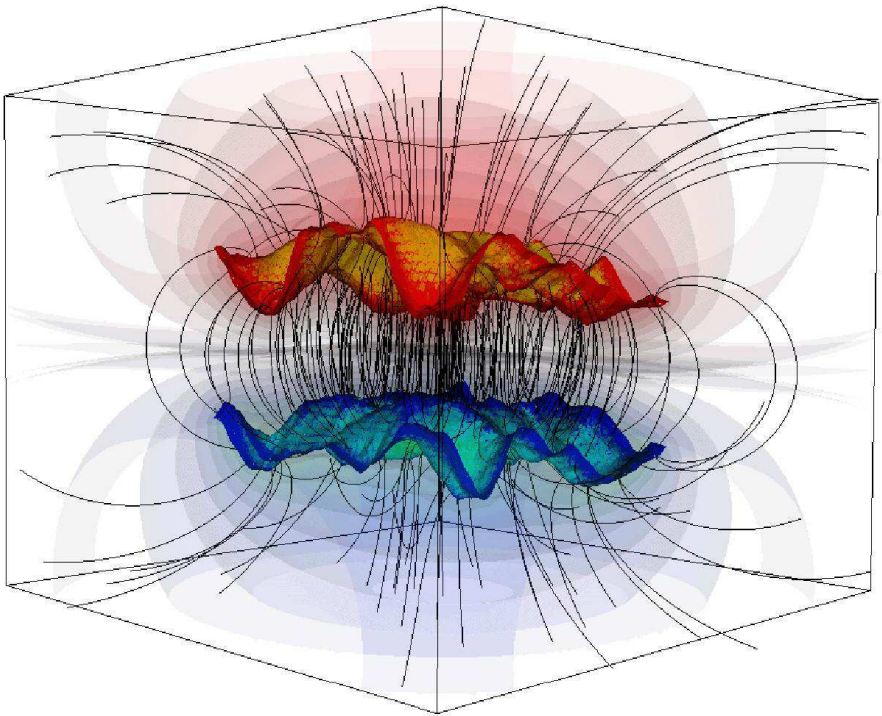}}}}
\caption{\label{fig:perlin} 
The surface charge density, the equipotential
surfaces and the lines of force for the system of two insulated randomly 
generated surfaces with charges $+Q$ and $-Q$. The random surfaces were generated using the Perlin noise algorithm contained in the VTK program package \cite{VTK}. The colours towards red
correspond to positive charge densities and the colours towards blue
correspond to negative charge densities.}
\end{figure}

%First we present the problem of the calculation of potential in the space
%produced when a charge is situated near a randomly generated surface. This
%problem is quite related to technological applications for problem with
%nonsmooth conducting surfaces. This example also shows that our method handles
%well situations in which conducting surfaces contain many details. In Fig.
%\ref{fig:random} is shown the charge distribution at the conducting plane
%produced by the placement of the point charge near it. {\bf Da li je ravnina na
%potencijalu ili uzemljena?}

\begin{figure}
\centerline{%\rotatebox{-90}
{\resizebox{1.0\textwidth}{!}
{\includegraphics{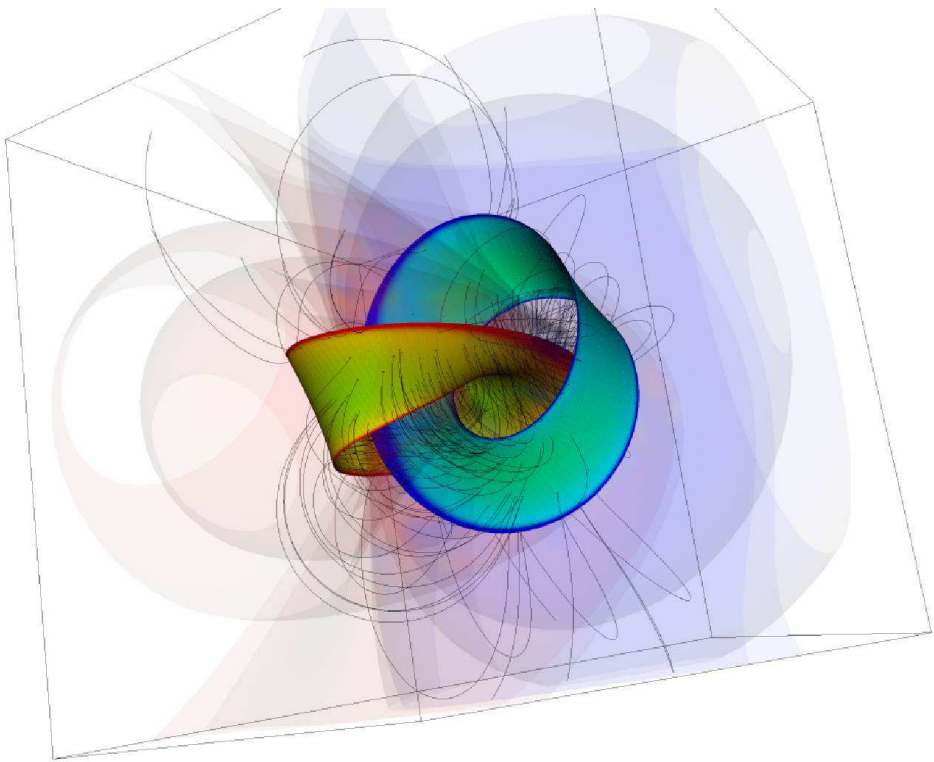}}}}
\caption{\label{fig:mobius} 
The surface charge density, the equipotential
surfaces and the lines of force for the system of two 
M\"{o}bius stripes with charges $+Q$ and $-Q$. The colours towards red
denote to positive charge densities and the colours towards blue
denote negative charge densities.}
\end{figure}

\begin{figure}
\centerline{%\rotatebox{-90}
{\resizebox{1.0\textwidth}{!}
{\includegraphics{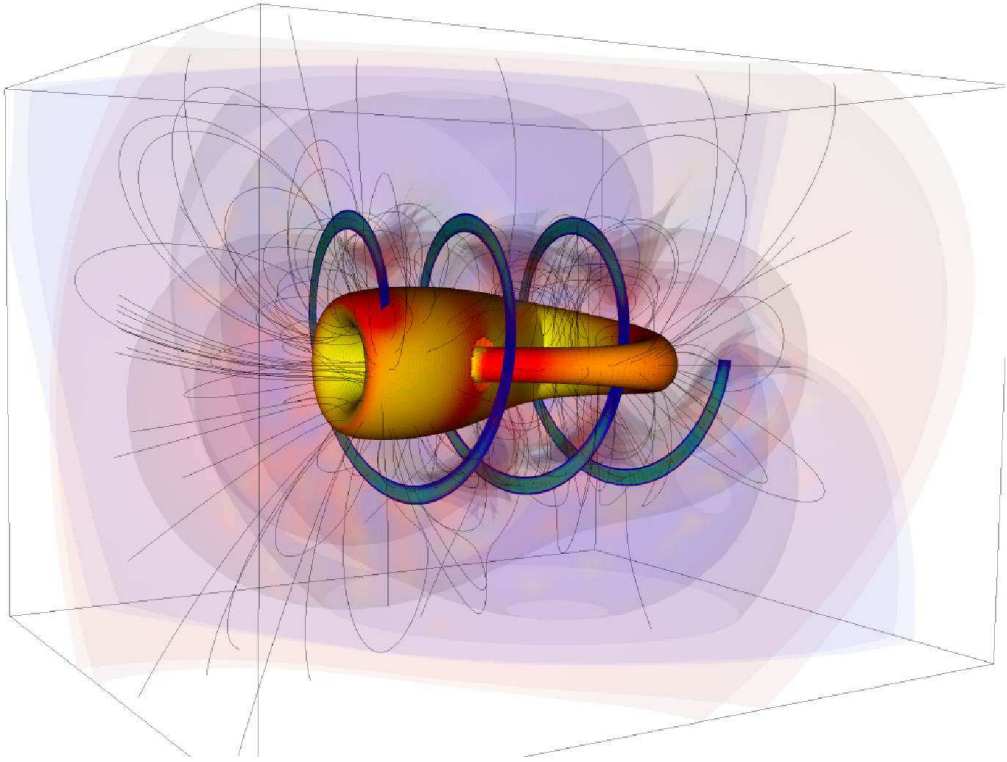}}}}
\caption{\label{fig:klein} 
The surface charge density, the equipotential
surfaces and the lines of force for the system of a segment of a spiral 
and a Klein bottle with charges $+Q$ and $-Q$. The colours towards red
represent positive charge densities while the colours towards blue
represent negative charge densities.}
\end{figure}

We study three different systems of which each consists of two insulated
conducting surfaces. One of the surfaces carries the charge $+Q$ while the
other carries the charge $-Q$. In this way, each of three studied systems
represents a capacitor. In all three cases, shown in Figs \ref{fig:perlin}, 
\ref{fig:mobius} and \ref{fig:klein}, the colours towards red stand for
positive charge densities while the colours towards blue represent negative
charge densities.

The first system consists of two randomly generated surfaces. The solution
in terms of the surface charge densities, equipotential surfaces and the lines of
force is depicted in Fig. \ref{fig:perlin}. The RH method handles without
difficulty randomly generated surfaces which confirms its suitability for
the treatment of geometries with high degree of variability or irregularity.
It is easy to extrapolate how the elaboration of this example might be used in
realistic systems with irregular surfaces. 

The second system consists of two interlocked M\"{o}bius stripes, as
displayed in Fig. \ref{fig:mobius}. By itself
it is a purely academic example, which, however in our case demonstrates the
effectiveness of the RH method in more complex topologies. Namely, M\"{o}bius stripe
has no inner and outer surface, as opposed to other surfaces considered
so far. Figure \ref{fig:mobius} shows the equilibrium surface charge
densities, the equipotential surfaces and the lines of force. 

The third system includes a segment of a spiral intertwined with a ``Klein
bottle". The purpose of this example is to show how far the geometric
complexity of the system can be pushed using the RH method. The equilibrium
situation for this system comprising the surface charge densities, equipotential 
surfaces and the lines of force is given in Fig. \ref{fig:klein}.

The examples given in this subsection demonstrate how RH method handles even
geometrically very complicated situations. The results presented
so far show that the RH method is a robust method which can be used in
treatment of very different geometries.  

\section{Technical characteristics of the method}

\label{tech}

In the preceding section we demonstrated the applicability of our method to a broad 
range of different configurations (arrangements of objects in space) 
and geometries (shapes of individual objects), charge distributions and 
potential boundary conditions,
as well as its reliability by comparisons with 
analytic solutions for selected problems. In this section we focus on
technical characteristics of the method.
% which reflect its efficiency in solving potential problems.

\begin{figure}
\centerline{%\rotatebox{-90}
{\resizebox{0.9\textwidth}{!}
{\includegraphics{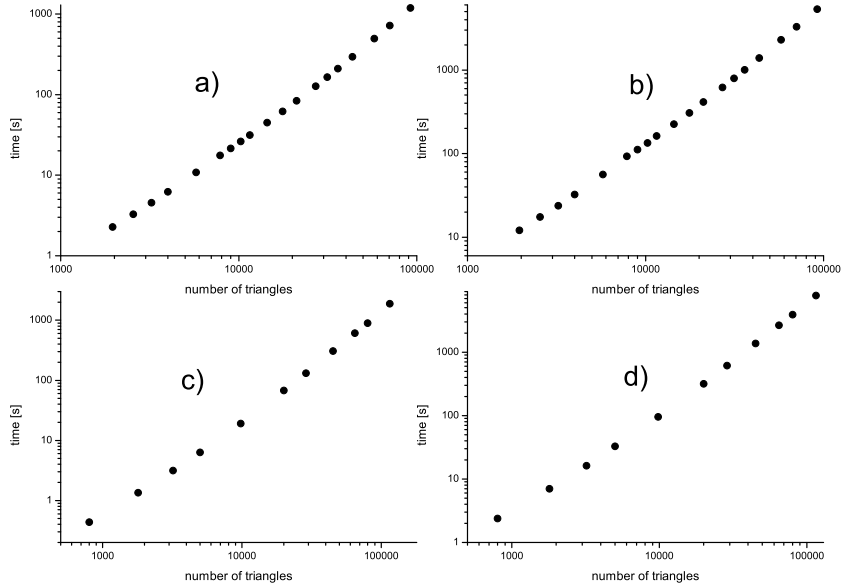}}}}
\caption{\label{fig:compl} 
The computational complexity of the RH method. a) The
dependence of the execution time of the first stage $t_{1}$ on the number of
triangles $N$ for the system of the conducting sphere at a fixed potential
and the point charge. b) The execution time of the second
stage $t_{2}$ as a function of $N$ for the same system as in a). c) The 
dependence of $t_{1}$ on $N$ for the system of a point charge and a plane at
a fixed potential. d) The dependence of $t_{2}$ on $N$ for the same system
as in c).}  
%The time of the execution of the algorithm in dependence on the number 
%of the surface elements. The circles represent the results of calculation %and
%the solid line is the result of the fit to the function of the form
%$t=C_{1} N^2 + C_{2} N \ln N$.}
\end{figure}

One of the most prominent characteristics of any algorithm is its {\em complexity}.
It specifies how the execution time of the algorithm changes depending on the 
number of elements that the algorithm deals with.
The complexity of our method determines how the time required to achieve the 
equipotentiality in the system scales with the number of surface elements $N$ 
for a specified geometry. To make a more thorough analysis of the 
computational complexity of the RH method, we chose to measure the dependence of 
the time of the execution of two stages in the algorithm of the RH method on 
the number of surface elements $N$. The first stage is the calculation of 
the initial potential at the surfaces of the considered system, while the 
second stage comprises successive iterations of the non-local transfer of 
the charge until equipotentiality is achieved. To account for possible 
dependence of the complexity on the geometry, we studied several 
geometries and configurations. For each total number of surface elements $N$, 
the initial division of the surfaces was used 
throughout the calculation, i.e. no iterative subdivisions of the surface 
elements described 
in section \ref{impl} were used.
Both the execution time of the first stage, $t_{1}$ and the 
execution time of the second stage, $t_{2}$, are well described 
by the power laws, i.e. $t_{i} \sim N^{\alpha_{i}}$, $i=1,2$. 
The first system in which the investigation of the complexity 
of the RH method was performed is the system of a point charge near a conducting sphere
at a fixed potential. 
%{\bf karakteristike}. 
In this system, the exponents are $\alpha_{1}=1.62$ and 
$\alpha_{2}=1.58$. For the system of the conducting plane kept at 
the exterior potential and a point charge and a similar system of the 
insulated plane and a point charge, we obtain very similar results. 
The exponents of the first stage are $\alpha_{1}=1.70$ and the exponents 
of the second stage amount to $\alpha_{2
}=1.64$ for both systems.
The log-log plots illustrating complexity for the specified systems are 
presented in Fig. \ref{fig:compl}.
The exponents $\alpha_{i}$ of the two stages are generally different, 
where $\alpha_{1}$ is slightly larger than $\alpha_{2}$ for all studied 
systems.   The results obtained show that the 
complexity is geometry dependent. The complexity difference of the two 
studied geometries is not large, but the general dependence of the 
complexity on geometry is an open question. Very similar results for the 
systems containing conducting planes indicate that there is no significant 
dependence of complexity on potential configuration. These results 
imply that RH method has no intrinsic complexity, but it is 
system dependent. However, a very important characteristic of the RH method 
emerges from the results of the studied systems. In all studied cases, 
the largest of the two exponents (which determines the complexity of the 
entire RH method) is considerably {\em less than 2} for $N$ of the order of 
$100$ to the order of $10^{5}$.
This feature, boosted by the easy parallelization and adaptive subdivision,
 makes the RH method competitive for large-scale 
problems involving a large number of surface elements.

%The investigation of the complexity of our method was performed 
%in the system of a charge near a conducting sphere {\bf karakteristike}.
%For each total number of surface elements $N$, the initial division of the %surfaces was used 
%throughout the calculation, i.e. no iterative subdivisions of the surface %elements described 
%in the section \ref{impl} were used. The execution time of our method was %determined for several 
%values $N$. The results are displayed in the form of a log-log graph in Fig. %\ref{fig:compl}.
%The best fit function of the obtained points has the form $f(N)=2.277 \,
%10^{-6} N^{2}+0.00208561 N \ln N$.
%The term $N \ln N$ represents the real complexity of our method, while the %term $N^2$ comes from the
%initial calculation of the potential at all $N$ surface elements. It is %worth emphasizing at this 
%point that this initialization of the potentials is the only time when $N^2$ %calculations need to be 
%done in a single iteration. At all subsequent iterations the number of %calculations scales as $N$.
%The crossover from the domination of the $N \ln N$ term to the domination of %the $N^2$ term 
%(when the contributions of these two terms are comparable) takes place at %the values of $N$ around 
%10000.

A significant advantage of the method introduced in this paper is connected
with {\em memory 
requirements}  for calculations with $N$ surface elements. Namely, the required memory scales 
{\em linearly with
$N$}, which is a tremendous reduction of the required memory capacitance compared
with, e.g. the existing BEM, where the 
required memory scales as $N^2$. 
%This characteristic makes our method a leading candidate for the
%treatment of very complex geometries involving many different scales. 
Even with a single processor, 
using our method it was possible to easily perform calculations with $N=2 \cdot
10^5$ surface elements \footnote{Using 1 GB of RAM it
could be possible to work with up to $10^7$ triangles.}. 

\begin{figure}
\centerline{%\rotatebox{-90}
{\resizebox{0.9\textwidth}{!}
{\includegraphics{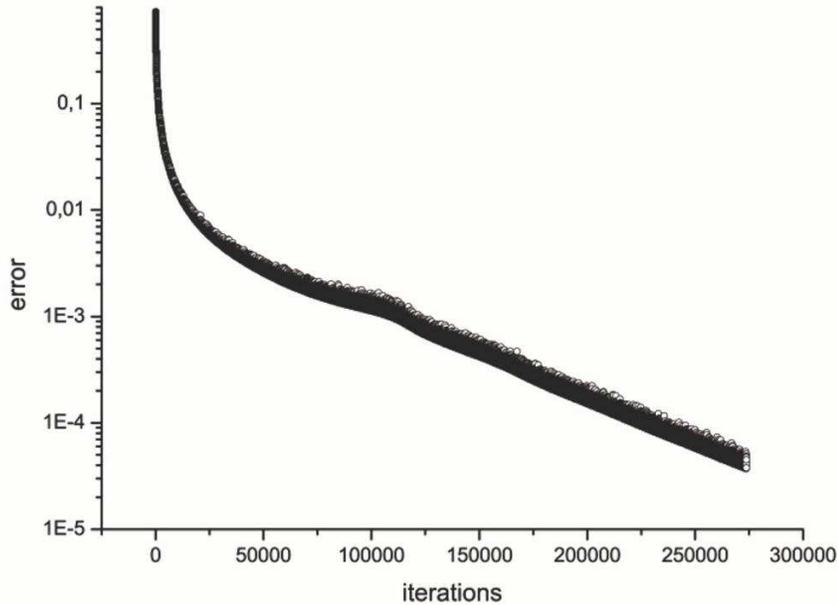}}}}
\caption{\label{fig:speed1} The speed of convergence for the system of the point charge close to the sphere kept at a fixed potential. 
The exponential character of the error reduction with the number of
iterations is evident. There is no effect of the Critical Slowing Down and the error
reduction efficiency is the same for any level of the error.}
\end{figure}

\begin{figure}
\centerline{%\rotatebox{-90}
{\resizebox{0.9\textwidth}{!}
{\includegraphics{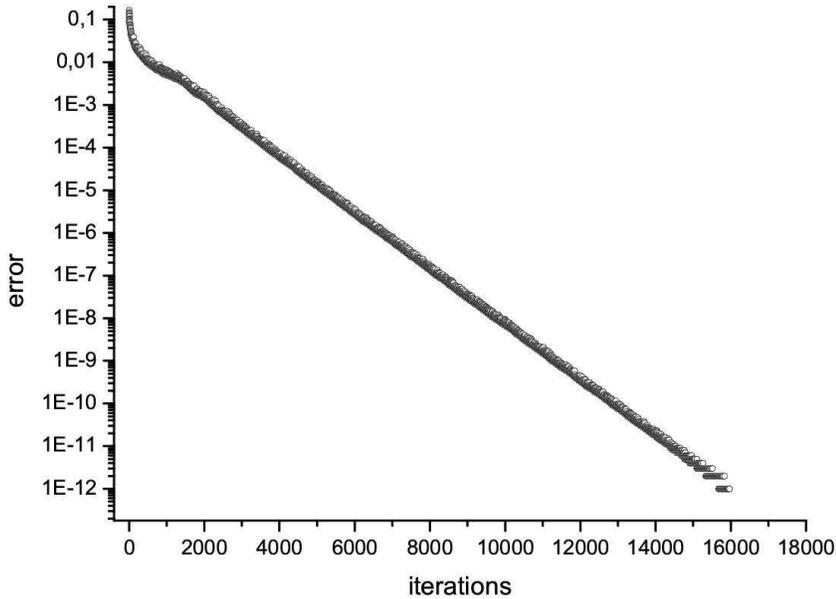}}}}
\caption{\label{fig:speed2} The speed of convergence for the system of the point charge located close to the
insulated conducting sphere. The error decreases exponentially with the number
of iterations for a broad range of error orders of magnitude. There is no effect
of the Critical Slowing Down.}
\end{figure}

\begin{figure}
\centerline{%\rotatebox{-90}
{\resizebox{0.9\textwidth}{!}
{\includegraphics{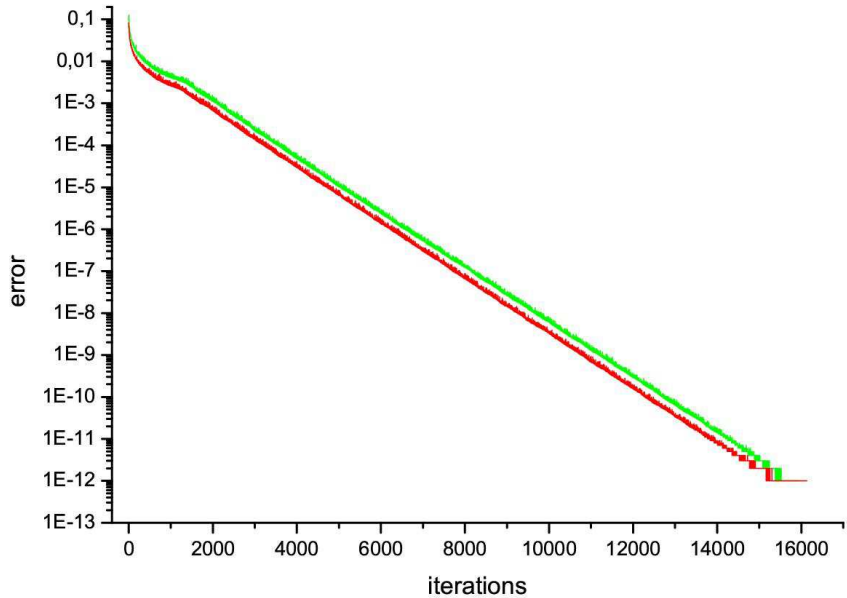}}}}
\caption{\label{fig:speed3} The speed of convergence for the system of a 
point charge and two insulated
conducting spheres (upper curve) compared with the speed of convergence of the
system discussed in Fig. \ref{fig:speed2} (lower curve). In both cases, the
error decreases exponentially with the number of iterations. The addition of one
more separate conducting surface does not deteriorate the speed of convergence.}
\end{figure}

\begin{figure}
\centerline{%\rotatebox{-90}
{\resizebox{0.9\textwidth}{!}
{\includegraphics{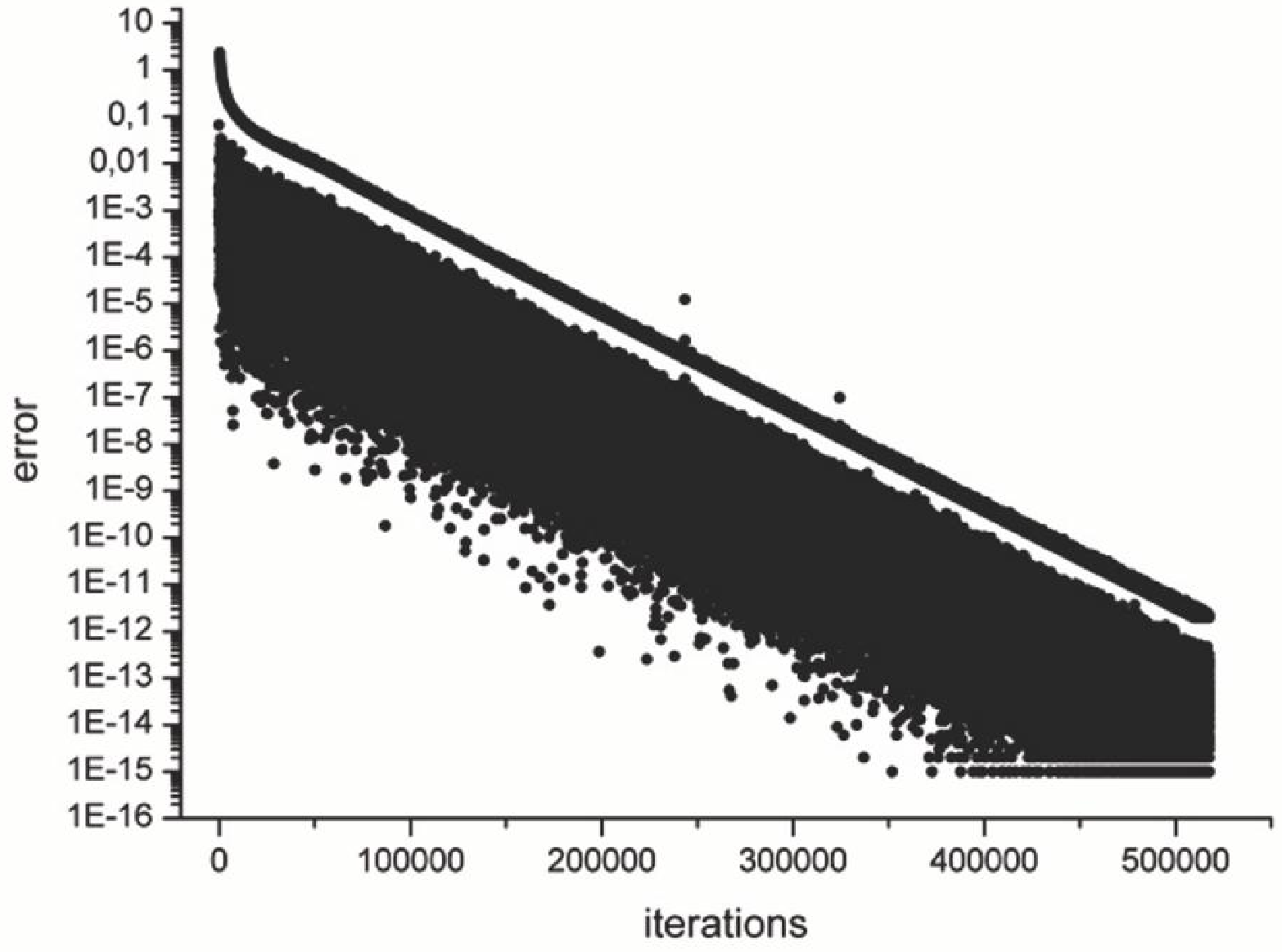}}}}
\caption{\label{fig:speed4} The speed of convergence for the system of a 
two randomly generated surfaces with charges of the equal amount and
opposite sign. The error decreases at worst
exponentially with the number of iterations without the CSD effect.}
\end{figure}

A further important advantage is  related to the speed of convergence of the method. The speed of 
convergence was studied in three different systems which all exhibit similar convergence properties.
The first system consists of a charge close to a sphere kept at a fixed potential.
% \footnote{The sphere was 
%actually maintained at the very small non-zero potential to avoid technical %complications.}. 
The behaviour of the 
error of the potential (defined as $(U_{\mathrm{maxdiff}}-U_{\mathrm{ext}})/U_{\mathrm{ext}}$ at a given
iteration, where $U_{\mathrm{ext}}$ is the external potential and $U_{\mathrm{maxdiff}}$ is
the potential differing the most from $U_{\mathrm{ext}}$) 
as a function of the number of iterations performed, which is quite
accurately equivalent to the execution time,
is given as a log-lin plot in Fig. \ref{fig:speed1}. From the 
figure it is clear that the error decreases exponentially with the number of iterations. An analogous graph 
for the system of the point charge close to an insulated conducting sphere is depicted in Fig. \ref{fig:speed2}.
The exponential dependence of the decrease of error on the number of iterations is evident. 
Finally,  we consider a system consisting of a point charge and two insulated conducting spheres.   
In this system each sphere ends up in a state with its own potential and charge redistribution 
is performed separately within each sphere. The error of the potential decays with the number 
of iterations in this case as well. Since we have two separate bodies on which
the potentials have 
to be equalized separately, it is reasonable to ask if the method converges 
more slowly than for the system
where the potential has to be equalized on a single surface only. To answer this question, in Fig. \ref{fig:speed3}
we present the dependence of the error of the potential on the number of iterations for the system with two 
spheres (the upper curve) and for the system with one sphere 
(the lower curve), which has already been
presented in Fig. \ref{fig:speed2}. This graph very clearly shows that the speed of convergence is the same 
for both systems. To complete the analysis of the speed of
convergence, we studied the system consisting
of two insulated conducting randomly generated surfaces carrying 
charges $+Q$ and $-Q$. The results obtained
are very similar to the preceding cases as shown in Fig. \ref{fig:speed4}.
Namely, for both surfaces the convergence is at worst exponential. The
larger dissipation of points in the graph reflects the random nature of the
studied geometry. 
The facts stated above indicate that our method handles systems of a different degree of 
geometric and configurational complexity with the same convergence speed, 
which is a very welcome property. It is important 
to state one additional convergence property common to 
Figs \ref{fig:speed1}, \ref{fig:speed2}, \ref{fig:speed3} and
\ref{fig:speed4}. 
The dependence of the error on the number of iterations is a straight line in the 
log-lin plot for a broad range of iteration numbers. In other words, no matter how small the error is, 
it is reduced by more or less the same percentage at subsequent iteration. The method has the same efficiency
of error reduction for all error sizes. Equivalently, the RH method equally
well reduces the error of the potential at the beginning of the calculation 
and at its end. 
Therefore, there is no effect of the {\em Critical Slowing Down} (CSD)  
present in some calculational methods \cite{CSD1,CSD2}, where the use of the
local information for the solution update leads to significant reduction in
the convergence speed with the number of iterations. This deficiency has
been overcome using the multigrid techniques \cite{multi} which, however,
require additional adjustment procedures. From the principles of the RH method
it is clear that the update of the potential is based on the global
information on the studied system. 
The absence of CSD could be explained by completely non-local charge transfers that always affect the
two worst points in the system making them the best points according to the criteria of equipotentiality.
One of the reasons of the very fast convergence is a natural adaptation of quantity of charge transfer being done 
at certain potential difference, i.e. at the reached precision. The RH method
handles the system equally efficiently at all levels of error.
In this sense, the system at a large potential difference of 
the order of 100 V (which represents a solution far from convergence) 
represents exactly the same problem for the method as the same system when it  
has reached differences of $10^{-20}$ V (already well converged
solution) and 
is being equally efficiently treated by the method.
The absence of the CSD, together with the $N$ scaling of the required memory, 
makes our method very suitable for large-scale high precision 
calculations.

Finally, we would like to address the technical features of the adaptive 
subdivision procedure described in section \ref{impl}. As already described, this
procedure achieves equipotentiality with the coarser division of the surface and
uses the charge distribution obtained as an initial distribution for the
calculations with the finer surface division in the next step. The open 
question at each step of this procedure is what proportion of triangles will be
subdivided and according to which criteria. Some of these criteria were proposed
in section \ref{impl}. We call the sequence of proportions and criteria for each step
{\em subdivision strategy}. With the subdivision strategy 
of dividing each right-angled triangle to four similar ones at each
subdivision step,
we achieved the reduction of the execution time around 25 \% for the system
of a point charge near a grounded plane. 
Although the RH method has excellent properties even without adaptive subdivision, the choice of the subdivision strategy is a matter of
the further optimization of the entire method.

\section{Applications beyond electrostatic problems}

\label{app}

The applicability of a numerical method to some physical problems depends on the
mathematical formulation of the problem. Many diverse physical problems
have an identical (or analogous) mathematical formulation and, therefore, can be
treated with the same numerical method. The RH method can be transferred from
the class of electrostatic problems to any physical problem with the same
mathematical formulation. An obvious application of the RH method beyond
electrostatics is thermostatics.

One possibility of extending the field of applicability of the RH method is its
transfer to the problems with identical mathematical 
formulation. However, except transferring the method in its original form, it is
possible to modify the RH method maintaining its basic principle of achieving
equipotentiality. In the modified RH method, the quantities such as the potential 
and the charge distribution have to be understood more generally. Let us
consider a problem in which it is necessary to find a spatial
configuration of the system in its physical state. The role of
the potential can be played by a quantity, let us call it {\em the generalized
potential}, which is spatially constant when the system is in the physical state.
The role of the charge distribution is then played by any {\em source}, 
spatial distribution of which has to be determined in order to equipotentialize
the generalized potential. These modifications of the RH method allow its
application in a very broad range of physical problems. In the following 
paragraph are given several illustrations of possible applications.

The (modified) RH method can be applied to semiclassical and quantum problems 
as well. Each
new application, apart from the modifications described above, requires some
technical adaptations. An example of the application of the RH method in
semiclassical problems is the treatment of the Thomas-Fermi approach. In this
case, the generalized potential becomes the energy of the system and the source
is the density of the particles. The generalized potential depends non-linearly
on the density owing to the approximation of the kinetic term. The modified RH
method handles this non-linear system equally well as the linear ones. Another
example of application is solving one-particle Schr\"{o}dinger equation in an
arbitrary potential. This is an eigenvalue problem which requires additional
adaptations of the original RH method. In the definition of the generalized
potential appear the terms with derivatives. In the discretized version of the
problem, these terms introduce non-locality into the problem. The ground 
state of the system can be obtained in a straightforward fashion \cite{mi2}. 
The problem of excited states and degeneracy is under current study \cite{mi2}.
Finally, there are indications that the modified RH method might be very useful
within the DFT approach \cite{multi,dacapo,KohnSham}. Recently, real space DFT methods have
emerged \cite{multi}. An implementation of real-space DFT based on the RH method is a promising approach for the treatment of electronic structure problems.

\section{Conclusions}

\label{concl}

In this paper we have presented a robust calculational method, 
which we have named the Robin Hood (RH) method, for solving a large
class of problems in electrostatics. The physical basis of the RH method is
the very intuitive concept of non-local transfer of the electric charge on the
conducting surfaces with the aim of achieving equipotentiality of all
conductors. The non-locality of the charge transfer and the long range of the
electrostatic interaction make possible the  global (i.e.
at the level of all conducting surfaces) improvement towards the
equipotentiality with each charge transfer.   
The method is characterized by many technical advantages which promote it into
an attractive tool for treating complex problems in electrostatics.
The memory requirements grow linearly with the number of surface
elements $N$. The complexity of the method itself scales as $N^{\alpha}$ with $\alpha < 2$ and is (somewhat) geometry dependent. The speed
of convergence is exponential and the Critical Slowing Down is not present,
which eliminates the necessity for procedures like multigrid techniques. 
Our method {\em does not} use multigrids \cite{Beck1,multi}. The results of the method are
``recyclable" in the sense that the charge distributions which satisfy the
condition of equipotentiality at coarse surface divisions can be
efficiently used as initial charge distributions for the calculations with finer
surface divisions. The adaptive subdivisions of selected surface elements can be
easily incorporated into the RH method, which can provide additional increase in
the computational efficiency of the method. 

Despite many appealing sides of the RH method demonstrated in this paper, there
are other issues to be investigated and possibly improved. Although the
method is already in many respects optimized,  the question of
performing charge transfers among more than two points at each conducting
surface, in linear problems such as electrostatics, remains. It is not excluded that for some $N_{\mathrm points} \ge 3$ the RH method delivers even better results, although a conceptually more complex approach. 
The strategy of the adaptive subdivisions of selected surface elements is
another aspect where careful optimization could lead to further improvements.
A more detailed study of the dependence of complexity on geometry might shed some light on geometries in which the method would provide an even better performance than for the systems studied in this paper.
These matters, along with other technical aspects, clearly constitute 
important questions worthy of further pursuit.

Besides rather academic examples shown in this paper, the RH method is suitable
for the electrostatic calculations of high precision in many practically 
very important examples such as the design of high energy
particle detectors, instruments for medical applications, 
circuit board design and many others.

Although the RH method is applicable to large classes of problems in electrostatics, it is worth investigating if the method could be modified to  encompass some other systems, such as those for which the explicit Green function is not available. 
The RH method can be easily transferred into the related fields of classical physics
and engineering where the mathematical background is equivalent. Furthermore, this
method provides a new way of handling quantum phenomena, the one-particle 
Schr\"{o}dinger equation and DFT being some of them. Although the strength of
the (modified) RH method in quantum problems is still to be properly gauged, it
is definitely a new alternative for unraveling the properties of important physical
systems.

%Advantages: 

%3D problem solved on a 2D surfaces (2D subspace) which is similar to
%BEM. 

%Possibility of use of the solution obtained in the previous iteration step -
%efficient 0.

%Robust method - works in cases where other methods fail (e.g. BEM)

%Requirement on computer 0 grows linearly with the number of POCs -
%possibility of attacking large scale problems. Large and detailed surfaces.
%0 of treating arbitrary surfaces, even random ones.

%Shortcomings:

%The knowledge of the analytic form of Green function is a prerequisite -
%problems with inhomogeneous media?

%Questions: 

%Applications to other technical problems or potential problems?

{\bf Acknowledgments.} The authors would like to thank dr. Radovan Brako and Prof. Dragan Poljak for useful
comments on the manuscript. We acknowledge the use of {\em VTK} and 
{\em Mathematica} program packages.
This work was supported by the Ministry of Science
and Technology of the Republic of Croatia under the contract 
numbers 0098001 and 0098002.

% main text
%\section{}
%\label{}

% \appendix

% \section{}
% \label{}

\end{document}